\documentstyle[epsfig,twocolumn,prl,aps]{revtex}
\begin{document}
\draft
\twocolumn[\hsize\textwidth\columnwidth\hsize\csname @twocolumnfalse\endcsname

\title{Reactive Hall response}
\author{X. Zotos$^1$, F. Naef$^1$, M. Long$^2$, 
and  P.Prelov\v sek$^{3,4}$ }

\address{ $^1$ Institut Romand de Recherche Num\'erique en Physique des
Mat\'eriaux (IRRMA), PPH-Ecublens, CH-1015 Lausanne, Switzerland }
\address{ $^2$ Department of Physics, University of Birmingham,
Edgbaston, Birmingham B15 2TT, United Kingdom} 
\address{ $^3$ Theoretische Physik, ETH-H\"onggerberg, 8093 Z\"urich, 
Switzerland}
\address{ $^4$ Faculty of Mathematics and Physics 
and J. Stefan Institute, 1001 Ljubljana, Slovenia }
\date{\today}
\maketitle
\begin{abstract}
The zero temperature Hall constant $R_H$, described by reactive 
(nondissipative) conductivities, is analyzed within linear response theory. 
It is found that in a certain limit, $R_H$ is directly related to the 
density dependence of the Drude weight implying a simple picture 
for the change of sign of charge carriers in the vicinity of a Mott-Hubbard 
transition. This novel formulation is applied to the calculation of $R_H$
in quasi-one dimensional and ladder prototype interacting electron systems.
\end{abstract}
\pacs{PACS numbers: 71.27.+a, 71.10.Fd, 72.15.Gd}
]

It is now well known that in strongly correlated systems the, 
zero temperature (T=0), reactive part of the conductivity can be used as a 
criterion of a metallic or insulating ground state\cite{kohn}. 
In particular, following the work of Kohn, the imaginary part of the 
conductivity, 
$\sigma''(\omega\rightarrow 0)=2D/\omega$, characterized by 
$D$ (now called the ``Drude weight" or charge stifness), 
can be related 
to the ground state energy density $\epsilon^0$ dependence on an applied 
fictitious flux $\phi$ as 
$D=(1/2)\partial^2 \epsilon^0/\partial \phi^2|_{\phi\rightarrow 0}$.  

A similar question is posed by the doping of an insulating state, where it 
would be interesting to have a simple description of the charge carriers 
sign as probed in a Hall experiment. 
For instance, we would like to describe the doping of a 
Mott-Hubbard insulator; within a semiclassical approach it is 
expected that the Hall 
constant $R_H\simeq +1/e\delta$, hole-like (positive) near 
half-filling ($\delta=1-n$, $n$=density), changing to 
$R_H\simeq -1/en$, electron-like at low densities, the turning point 
depending on the interaction. 

Over the recent years, ingenious ways have been proposed\cite{cb,rkc} for 
characterizing this sign change and strongly correlated electron systems, 
as the $t-J$ model have been studied.
In particular, following the suggestion to focus at the T=0 
Hall constant within linear response theory\cite{pp},   
the $R_H$ of a hole in the $t-J$ model was analyzed and a numerical 
method was proposed for calculating the Hall response in 
ladder systems\cite{plmz}.
This activity is partly motivated by the physics of high 
temperature superconductors viewed as doped Mott-Hubbard insulators and 
related Hall measurements showing a change of the sign of carriers 
with doping\cite{ong}.

In this work, we will show that within a certain frequency-$\omega$, 
wavevector-$q$ limiting procedure, the T=0, 
$\omega\rightarrow 0$, thus ``reactive" Hall constant, is simply related 
to the density dependence of the Drude weight. Following this point of view,  
we recover in a straightforward way: (i) the semiclassical expressions for 
$R_H$ at low density and near an insulating state, (ii) a physical picture of 
the sign change of carriers in the vicinity of a 
Mott-Hubbard transition and its dependence on interaction strength, 
(iii) a common expression used to describe the Hall constant in quasi-one 
dimensional conductors described by a band picture\cite{maki}, 
(iv) good accord with $R_H$ for ladder systems calculated using the 
numerical method proposed in \cite{plmz}.

\vspace{0.3cm}
{\it The Hamiltonian}~~In the following we will consider a generic 
Hamiltonian for fermions on a lattice, where for simplicity we describe the 
kinetic energy term by a one band tight binding model; 
it is straightforward to extend this formulation to a many-band or 
continuum system.
The sites are labeled $l (m)$ along the $x (y)$-direction
with periodic boundary conditions in both directions:

\begin{eqnarray}
H&=&(-t)\sum_{l,m} e^{i\phi^x(t)} e^{iA_m} c^{\dagger}_{l+1,m}c_{l,m}+h.c.
\nonumber\\
&+&(-t')\sum_{l,m} e^{i\phi^y_{m+1/2}(t)} c^{\dagger}_{l,m+1}c_{l,m}+h.c.
\nonumber\\
&+&\hat U,~~~~~~l=1,...,L_x;~~m=1,...,L_y.
\label{ham}
\end{eqnarray}

\noindent
$c_{l,m}(c^{\dagger}_{l,m})$ is an annihilation (creation) operator 
at site $(l,m)$ and the spin is neglected as it enters in a trivial way in the 
formulation. The $\hat U$ term can represent a many-particle interaction 
or a one particle potential. We take the lattice constant so as to 
consider a unit volume, electric charge $e=1$ and
$\hbar=1$. We add a magnetic field along 
the $z$-direction, modulated by a one component wavevector-$q$ 
along the $y$-direction, generated by the vector potential $A_m$; 
this allows to take the zero magnetic field limit smoothly\cite{fuku}:
\begin{eqnarray}
A_m&=&e^{iqm}\frac{iB}{2\sin(q/2)}\simeq e^{iqm}\frac{iB}{q}
\nonumber\\
B_{m+1/2}&=&-(A_{m+1}-A_m)=B e^{iq(m+1/2)}
\label{fieldb}
\end{eqnarray}

\noindent
(for convenience, we will present the long wavelength limit, 
substituting $2\sin(q/2)\rightarrow q $).
Electric fields along the $x,y$ directions are generated by time dependent 
vector potentials:

\begin{eqnarray}
\phi^{x,y}(t)&=&\frac{E^{x,y}(t)}{iz},~~
\phi^y_{m+1/2}(t)=e^{iq(m+1/2)}\phi^y(t);
\nonumber\\
E^{x}(t)&=&E^x e^{-izt},~~E^{y}(t)=iE^y e^{-izt};~~~z=\omega+i\eta~.
\label{phixy}
\end{eqnarray}

\noindent
Currents are defined through derivatives of the Hamiltonian expanded 
to second order in $\phi^{x,y}$: 
\begin{eqnarray}
J^x=-\frac{\partial H}{\partial \phi^x},~~~~~~
J^y_q=-\frac{\partial H}{\partial \phi^y}~,
\label{currents}
\end{eqnarray}

\noindent
with the paramagnetic parts:
\begin{eqnarray}
j^x&=&t\sum_{l,m}(ie^{iA_m} c^{\dagger}_{l+1,m}c_{l,m}+h.c.)
\nonumber\\
j^y_{q}&=& t'\sum_{l,m} e^{iq(m+1/2)}
(i c^{\dagger}_{l,m+1}c_{l,m}+h.c.).
\label{jpar}
\end{eqnarray}

\vspace{0.3cm}
{\it The reactive Hall response}~~From standard linear response theory 
we obtain:  

\begin{eqnarray}
\langle J^x\rangle&=&\sigma_{j^x j^x}E^x(t)+\sigma_{j^x j^y_q}E^y(t)
\nonumber\\
\langle J^y_q\rangle&=&\sigma_{j^y_q j^x}E^x(t)+\sigma_{j^y_q j^y_q}E^y(t)~.
\label{respjj}
\end{eqnarray}

\noindent
$\langle...\rangle$ are ground state expectation values in the 
presence of the magnetic field, with the conductivities  

\begin{eqnarray}
\sigma_{j^{\alpha}j^{\beta}}&=&\frac{i}{z}
(\langle\frac{\partial^2H}{\partial\phi^{\alpha}\partial\phi^{\beta}}\rangle-
\chi_{j^{\alpha}j^{\beta}}),
\nonumber\\
\chi_{AB}&=&i\int^{\infty}_0 dt e^{izt} \langle [A(t),B] \rangle.
\label{sigma}
\end{eqnarray}

\noindent
Now, in contrast to the usual derivation of the Hall constant expression, 
we will keep the $q-$dependence explicit by converting the 
current-current to current-density correlations using the continuity equation:

\begin{eqnarray}
\langle J^x\rangle&=&\sigma_{j^x j^x}E^x(t)+\frac{1}{q}\chi_{j^x n_q}E^y(t)
\nonumber\\
\langle J^y_q\rangle&=&-\frac{1}{q}\chi_{n_q j^x}E^x(t)+
(\frac{z}{q})^2\chi_{n_q n_q}\frac{i}{z}E^y(t)~,
\label{respjn}
\end{eqnarray}

\noindent
with $n_q=\sum_{l,m} (-ie^{iqm}) c^{\dagger}_{l,m}c_{l,m}$.

At T=0, the response is non-dissipative so we will study 
the reactive (out-of phase) induced currents. 
Furthermore, at this point we will consider the ``screening" (or slow)
response in the $y-$direction, by taking the $(q,\omega)$ limits in 
the order $\omega\rightarrow 0$ first and 
$q\rightarrow 0$ last; in the usual ``transport" (or fast) response  
the limits are in the opposite order\cite{lutt}.
As we will discuss below, this approach leads to a simple physical 
picture for the Hall constant and it might be argued that at least for certain 
cases, for example for a system of finite size in the $y-$direction, 
it is indeed the right one. The expressions (\ref{respjn}) for the 
currents become:

\begin{eqnarray}
\langle J^x\rangle_0&=&\sigma_{j^x j^x}''(\omega\rightarrow 0))(iE^x(t))
\nonumber\\
&+&\frac{1}{q}\chi_{j^x n_q}'(\omega=0)E^y(t)
\nonumber\\
\langle J^y_q\rangle_0 &=&-\frac{1}{q}\chi_{n_q j^x}'(\omega=0)E^x(t)
\nonumber\\
&+&(\frac{\omega}{q})^2\frac{1}{\omega}\chi_{n_q n_q}'(\omega=0)(iE^y(t))~,
\label{inphase}
\end{eqnarray}

\noindent
where the subscript zero denotes the leading order in $\omega$ response, 
\begin{eqnarray}
\chi_{AB}'(\omega=0)=\sum_{n>0}
\frac{\langle 0|A|n\rangle\langle n|B|0\rangle+h.c.}{E_n-E_0},
\label{chiab}
\end{eqnarray}

\noindent
and $|n\rangle (E_n)$ are eigenstates (eigenvalues) of the Hamiltonian 
in the presence of the magnetic field.

Now, following Kohn's observation\cite{kohn}, we can identify the 
different terms as derivatives of the ground state energy density 
$\epsilon^0$ of a 
fictitious Hamiltonian depending on static $\phi^x,\mu_q$ fields:

\begin{eqnarray}
H&=&(-t)\sum_{l,m} (e^{i\phi^x} e^{iA_m} c^{\dagger}_{l+1,m}c_{l,m}+h.c.)
\nonumber\\
&+&(-t')\sum_{l,m} (c^{\dagger}_{l,m+1}c_{l,m}+h.c.)+\mu_q n_q +\hat U.
\label{ficth}
\end{eqnarray}

\noindent
For $H(\lambda,\mu)$, using the following identity,

\begin{eqnarray}
\epsilon^0_{\mu\lambda}=\frac{\partial^2 \epsilon^0}
{\partial\mu \partial \lambda}&=&
\langle 0|\frac{\partial^2 H}{\partial\mu \partial\lambda}|0\rangle
\nonumber\\
&-&\sum_{m>0}\frac{\langle 0|\frac{\partial H}{\partial \mu}|m\rangle
\langle m|\frac{\partial H}{\partial \lambda}|0\rangle+h.c.}
{E_m-E_0}~,
\label{ident}
\end{eqnarray}

\noindent
we can rewrite the currents as:

\begin{eqnarray}
\langle J^x\rangle_0&=&
\frac{\epsilon^0_{\phi^x\phi^x}}{\omega}(iE^x(t))+
(\frac{-1}{q})\epsilon^0_{\phi^x\mu_q}E^y(t)
\nonumber\\
\langle J^y_q\rangle_0&=&
\frac{1}{q}\epsilon^0_{\mu_q\phi^x}E^x(t)
-\frac{\omega}{q^2}\epsilon^0_{\mu_q\mu_q}(iE^y(t)).
\label{fictj}
\end{eqnarray}

\noindent
Finally, setting $\langle J^y_q\rangle_0=0$ we determine 
the ``reactive" Hall constant:

\begin{eqnarray}
R_H\equiv -\frac{1}{B} \frac{E^y}{\langle J^x\rangle_0}=
(-\frac{q}{B}) \frac{\epsilon^0_{\mu_q\phi^x}}
{\epsilon^0_{\phi^x\phi^x} \epsilon^0_{\mu_q\mu_q} + 
\epsilon^0_{\mu_q\phi^x} \epsilon^0_{\phi^x\mu_q}}~.
\end{eqnarray}

\noindent
Neglecting the cross-terms $\epsilon^0_{\mu_q\phi^x} \epsilon^0_{\phi^x\mu_q}$
and Taylor expanding the numerator in $B$, we can rewrite $R_H$ as:

\begin{eqnarray}
R_H=q \frac{\frac{\partial^3 \epsilon^0}
{\partial B\partial \mu_q\partial \phi^x}}
{\epsilon^0_{\phi^x\phi^x} \epsilon^0_{\mu_q\mu_q}}=
q \frac{
\frac{\partial}{\partial\mu_q}
(\frac{\partial^2 \epsilon^0}{\partial B\partial \phi^x})}
{\epsilon^0_{\phi^x\phi^x} \epsilon^0_{\mu_q\mu_q}}~.
\end{eqnarray}

\noindent
Using (\ref{ident}) we find the final expression:
\begin{eqnarray}
R_H=-\frac{\frac{\partial D_q}{\partial \mu_q}}{D \kappa_q}
\label{rhdq}
\end{eqnarray}

\noindent
where,
\begin{eqnarray}
D_q&=&\frac{1}{2}\lbrack \langle 0|-T^x_q|0\rangle
\nonumber\\
&-&\sum_m\frac{\langle 0|j^x|m\rangle
\langle m|j^x_q|0\rangle+h.c.}{\epsilon_m-\epsilon_0}\rbrack,
\nonumber\\
j^x_q&=&(-t)\sum_{l,m} (-ie^{iqm})(i c^{\dagger}_{l+1,m}c_{l,m}+h.c.), 
\nonumber\\
T^x_q&=&(-t)\sum_{l,m} (-ie^{iqm})(c^{\dagger}_{l+1,m}c_{l,m}+h.c.). 
\label{dq}
\end{eqnarray}

\noindent 
$D=\frac{1}{2}\epsilon^0_{\phi^x\phi^x}$, the Drude weight, is identical 
to $D_q$ by the replacement of $j^x_q$ ($T^x_q$) 
by $j^x$ ($T^x$). 
$\kappa_q=\epsilon^0_{\mu_q\mu_q}=\partial n_q/\partial \mu_q$ is the 
compressibility corresponding to the density modulation $n_q$.
Notice that the spatial dependence of $j^x_q$ and $n_q$ is the 
same as that of $A_m$. 

Taking the $q\rightarrow 0$ limit, we obtain a particularly simple 
expression for $R_H$:

\begin{eqnarray}
R_H=-\frac{1}{D}\frac{\partial D}{\partial n}.
\label{naive}
\end{eqnarray}

\noindent
A handwaving argument leading to expression (\ref{naive})     
for $t'\to 0$ is as follows: $A_m$ corresponds to a twist of
boundary conditions on chain$-m$, inducing an extra current on each chain
proportional to $D$ (besides the uniform one induced by the flux $\phi^x$);
minimization of the energy at fixed $x-$current gives rise to an 
$m-$dependent charge density. This induced charge density can 
then be canceled by the ``Hall potential" $\mu_q$\cite{plmz}.
Note that a similar idea, analyzing the Hall constant in terms of 
independent channels (edge states), exists in the literature of the 
Quantum Hall effect\cite{qhe}.

This expression is appealing as it gives a direct, intuitive 
understanding for the change of sign of charge carriers in the vicinity 
of a metal-insulator transition. 
First, at low densities, $D\propto n$ giving $R_H\simeq -1/n$;  
close to a Mott insulator $D\propto \delta=1-n$, 
implying $R_H\simeq +1/\delta$. Furthermore, we obtain a change of sign in the 
vicinity of a Mott transition at a density which depends on the interaction 
strength and is given by the position of the maximum of $D$. 
Second, for independent electrons, where $D$ is proportional to the 
kinetic energy, by taking the limit $t'\rightarrow 0$  
and calculating $D$ as a sum of $D$'s for
individual $x-$chains, we obtain from (\ref{naive}):
\begin{equation}
D=\frac{2t}{\pi}\sin(\frac{\pi n}{2}),~~~
R_H=-\frac{\pi}{2}\frac{1}{\tan(\frac{\pi n}{2})},
\label{1d}
\end{equation}

\noindent 
an expression used for the Hall constant of quasi one-dimensional 
compounds\cite{maki}. 
Considering that the $t'\rightarrow 0$ limit might by subtle, 
it is of particular theoretical and experimental interest  
whether the Hall constant of quasi-one dimensional correlated 
systems \cite{str} is indeed given by the expression and thus related to the 
Drude weight of the individual chains. The same applies for the 
transverse Hall effect of weakly coupled planes. 

\vspace{0.3cm}
{\it Examples}~~In this section we present a generic picture 
for the behavior of the Hall constant for models of strongly 
correlated fermions showing a Mott-Hubbard metal-insulator transition. 
This picture emerges, on the one hand, by an exact 
calculation of $R_H$ for ladder systems using the numerical method of 
ref.\cite{plmz} and on the other hand, from the expression (\ref{naive}) 
assuming nearly decoupled chains ($t'\rightarrow 0$)   
and calculating $D(n)$ for each chain analytically using the Bethe ansatz 
method\cite{hald,ky}. It is clear that this analytical approach refers 
to either ladder (with $t'\rightarrow 0$) or quasi one-dimensional models. 

Three prototype models will be discussed: 
the Hubbard model, as the most experimentally relevant, the spinless 
fermions model (``t-V") showing both a metallic and an insulating phase 
depending on interaction strength and the supersymmetric $t-J$ model. 

(i) The {\it Hubbard model} is given by the Hamiltonian:

\begin{eqnarray}
&H&=(-t)\sum_{l,m}(c^{\dagger}_{l+1,m,\sigma}c_{l,m,\sigma}+h.c.)
\nonumber\\
&+&(-t')\sum_{l,m}(c^{\dagger}_{l,m+1,\sigma}c_{l,m,\sigma}+h.c.)
+U\sum_{l,m} n_{l,m,\uparrow}n_{l,m,\downarrow}~.
\label{hub}
\end{eqnarray}

\noindent
$c_{l,m,\sigma}(c^{\dagger}_{l,m,\sigma})$ is an annihilation (creation) 
operator at site $(l,m)$ of a fermion with spin $\sigma=\uparrow,\downarrow$.
$R_H$ extracted from a Bethe ansatz calculation of $D(n)$ for the 
one dimensional Hubbard model\cite{ky} is shown in Fig.~1.

\begin{figure}[h]
\epsfxsize=2.5 in\centerline{\epsffile{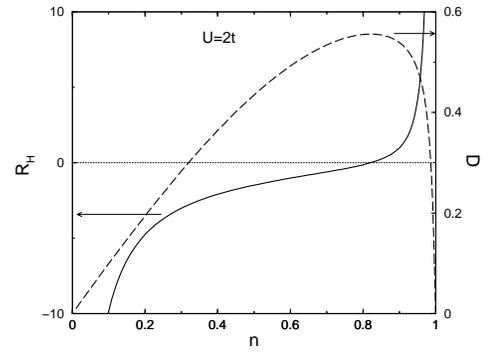}}
\caption{$R_H$ for the Hubbard model from expression (\ref{naive}) 
for $t'\rightarrow 0$.}
\label{fig1}
\end{figure}

This behavior is characteristic of correlated systems  
undergoing a metal-insulator transition at half-filling: at low 
densities $R_H\simeq -1/n$, while near half-filling $R_H\simeq +1/\delta$, 
the position of change of sign of the carriers depending on the details of the 
interaction.

(ii) The {\it t-V} model on a ladder is given by:

\begin{eqnarray}
H&=&(-t)\sum_{l,m}(c^{\dagger}_{l+1,m}c_{l,m}+h.c.)
\nonumber\\
&+&(-t')\sum_{l}(c^{\dagger}_{l,1}c_{l,2}+h.c.)+V\sum_{l,m} n_{l,m}n_{l+1,m}.
\label{tv}
\end{eqnarray}

\noindent
Here and in the following $l=1,...,L_x, m=1,2$. 
For a single chain, this model describes a metallic phase at all 
densities for $V< 2t$, while for $V> 2t$ it is an insulator at half-filling. 
In Fig.~2 we show $R_H$ calculated numerically on 
finite systems for two values of $t'$ and  
analytically from (\ref{naive}) in the $t'\rightarrow 0$ limit. 
The numerical evaluation being especially sensitive to 
finite size effects for $t'\rightarrow 0$, 
we study relatively large values of $t'$.

Results for $R_H$ clearly show the difference between the
metallic regime $V=t$, where at half-filling ($n=0.5$) we get
$R_H=0$, while in the insulating regime $V=4t$, we are
dealing with $R_H(n\to 0.5) \to \infty$.

\begin{figure}[h]
\epsfxsize=7.5 cm\centerline{\epsffile{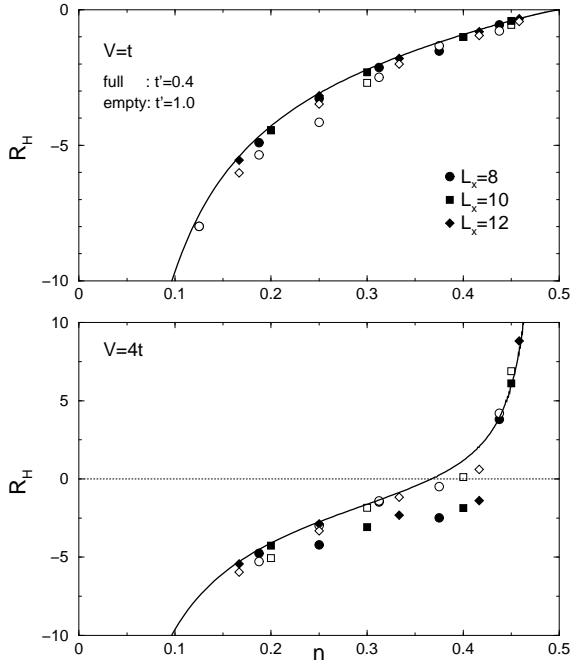}}
\caption{$R_H$ for the $t-V$ ladder from expression (\ref{naive}) 
for $t'\rightarrow 0$ (continuous line) and from a numerical evaluation 
(symbols). $V=t (4t)$, metallic (insulating) phase at $n=0.5$.}
\label{fig2}
\end{figure}

(iii) The {\it t-J} model on a ladder is given by the Hamiltonian:

\begin{eqnarray}
H&=&(-t)\sum_{l,m}(c^{\dagger}_{l+1,m,\sigma}c_{l,m,\sigma}+h.c.)
\nonumber\\
&+&(-t')\sum_{l}(c^{\dagger}_{l,1,\sigma}c_{l,2,\sigma}+h.c.)
\nonumber\\
&+&J\sum_{l,m} (\vec S_{l,m}\vec S_{l+1,m}-\frac{1}{4}n_{l,m}n_{l+1,m})~.
\label{tj}
\end{eqnarray}

\noindent
$\vec S_{lm}$ is the spin operator at site $(l,m)$ and
the double occupancy on a site is forbidden.

\begin{figure}[h]
\epsfxsize=2.5 in\centerline{\epsffile{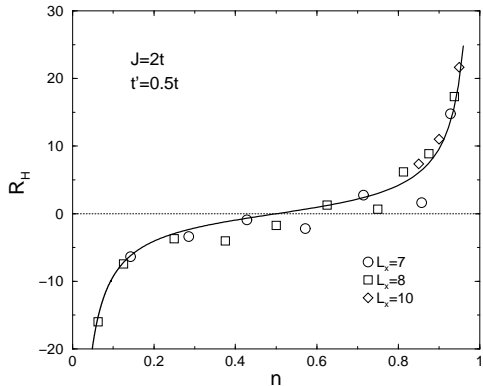}}
\caption{$R_H$ for the $t-J$ ladder from expression (\ref{naive})
for $t'\rightarrow 0$ (continuous line) and from a numerical evaluation 
(symbols).}
\label{fig3}
\end{figure}

In Fig.~3 we show again $R_H$ calculated analytically 
for the ``supersymmetric" model, $J=2t$, and by numerical 
evaluation for $t'=0.5t$ and different size systems.

The above three examples show a remarkable agreement between the 
numerical evaluation of $R_H$ on finite size systems 
using the method of ref.\cite{plmz} (at finite $t'$) and the analytical 
calculation using (\ref{naive}) for $t'\rightarrow 0$,  
indicating a relative insensitivity on the transverse coupling $t'$ 
for ladders.
These results confirm the intuitive picture for the behavior of the Hall 
constant in the vicinity of a metal-insulator transition and present 
an intriguing link between the Hall constant and the Drude weight. 
It is possible that $R_H$ is dominated at low temperatures by correlations 
and not the relaxation mechanism so this formulation could have more 
general validity. 

In conclusion, the emerging simple physical picture raises the question
of the relation of this novel formulation to the traditional semiclassical 
approach to the Hall constant, its range of validity, the role of 
relaxation in the description of the Hall effect
and of the perspectives for an extension at finite temperatures.

Part of this work was done during visits of (P.P.) and (M.L.) at IRRMA as 
academic guests of EPFL.
X.Z. and F.N. acknowledge support by the Swiss National Foundation 
grant No. 20-49486.96, the EPFL, the Univ. of Fribourg and the Univ. of 
Neuch\^atel.

\end{document}